%% file: arXiv-ainsworth-jcdl2011-long.tex
\documentclass{acm_proc_article-sp-tweeked}
\usepackage{amsmath}
\usepackage{listings}
\usepackage{graphicx}
\usepackage{subfigure}
\usepackage[usenames]{color}
\usepackage[letterpaper]{hyperref}

\begin{document}

\markboth{S. Ainsworth et al.}{How Much of the Web Is Archived?}
\pagenumbering{arabic}


\title{How Much of the Web Is Archived?}

\numberofauthors{1}%
\author{
\alignauthor
Scott G. Ainsworth,
Ahmed AlSum,
Hany SalahEldeen,\\
Michele C. Weigle,
Michael L. Nelson\\
	\affaddr{Old Dominion University}
	\affaddr{Norfolk, VA, USA}
	\email{\{sainswor, aalsum, hany, mweigle, mln\}@cs.odu.edu}
}
\maketitle


\begin{abstract}
Although the Internet Archive's Wayback Machine is the largest and
most well-known web archive, there have been a number of public web
archives that have emerged in the last several years.  With varying
resources, audiences and collection development policies, these
archives have varying levels of overlap with each other.  While
individual archives can be measured in terms of number of URIs,
number of copies per URI, and intersection with other archives, to
date there has been no answer to the question ``How much of the Web
is archived?'' We study the question by approximating the Web using sample
URIs from DMOZ, Delicious, Bitly, and search engine indexes; and,
counting the number of copies of the sample URIs exist in various public
web archives.  Each sample set provides its own bias. The results from our sample sets indicate that range from 35\%-90\% of the Web has at least one archived
copy, 17\%-49\% has between 2-5 copies, 1\%-8\% has 6-10 copies, and 8\%-63\% has
more than 10 copies in public web archives.  The number of URI
copies varies as a function of time, but no more than 31.3\% of URIs
are archived more than once per month.
\end{abstract}






\input{Introduction}

\input{RelatedWork}

\input{Experiment-Overview}

\input{Experiment-SampleSelection-intro}
\input{Experiment-SampleSelection-dmoz}

\input{Experiment-SampleSelection-delicious}
\input{Experiment-SampleSelection-bitly}
\input{Experiment-SampleSelection-searchengine}
\input{Experiment-CurrentState}
\input{Experiment-MementoDiscovery}
\input{Experiment-AgeEstimation}

\input{Results}

\input{Analysis}
\input{Conclusion}


\section{Acknowledgments}
This work is supported in part by the Library of Congress. We would like to thank Herbert Van de Sompel and Robert Sanderson from Los Alamos National Laboratory, and Kris Carpenter Negulescu and Bradley Tofel from Internet Archive for their positive comments and explanations.


\bibliographystyle{plain}
\bibliography{references}

\end{document}

%% file: Introduction.tex
%


\section{Introduction}

\newcommand{\squishlist}{
  \begin{list}{$\bullet$}
  {
   \setlength{\itemsep}{0pt}
   \setlength{\parsep}{3pt}
   \setlength{\topsep}{3pt}
   \setlength{\partopsep}{0pt}
   \setlength{\leftmargin}{1.5em}
   \setlength{\labelwidth}{1em}
   \setlength{\labelsep}{0.5em} 
   } 
 }

\newcommand{\squishlisttwo}{
  \begin{list}{$\bullet$}
  {
   \setlength{\itemsep}{0pt}
   \setlength{\parsep}{0pt}
   \setlength{\topsep}{0pt}
   \setlength{\partopsep}{0pt}
   \setlength{\leftmargin}{2em}
   \setlength{\labelwidth}{1.5em}
   \setlength{\labelsep}{0.5em} } }

\newcommand{\squishend}{
   \end{list}  }

With more and more of our business, academic, and cultural discourse
contained primarily or exclusively on the web, there has been an
increased attention to the problem of web archiving.  The focal
point of such discussions is the Internet Archive's Wayback Machine,
which began archiving the web in 1996 and as of 2010 had over 1.5
billion unique URIs \cite{carpenter-ndiipp10}, thus making it the
largest, longest-running and most well known of publicly available
web archives.  On the other hand, given the increased attention to
the problem of digital preservation, the proliferation of production
quality web crawling tools,  and the falling cost of resources
required for preservation (e.g., storage and bandwidth), there has
been a proliferation of additional public web archives at universities,
national libraries, and other organizations.  They differ in a
variety of ways, including scale, ingest models, collection development
policies, software employed for crawling and archiving.  

Anecdotally we know that these additional archives have some degree
of overlap with the Internet Archive's Wayback Machine, but at
the same time they are also not proper subsets of the Wayback
Machine.  This leads to a common question often asked of those working in
digital preservation: ``How much of the Web is archived?''
We are unaware of any prior attempt to address this question.  
We do know the approximate size of the Wayback Machine, and 
we do know that Google has measured the web to contain at least one
trillion unique URIs (though they do not claim to index all of
them)\footnote{\url{http://googleblog.blogspot.com/2008/07/we-knew-web-was-big.html}}.

In this paper, we do not attempt to measure the absolute size of
various public web archives.  Instead, we sample URIs from a variety
of sources (DMOZ, Delicious, Bitly, and search engine indices) and
report on the number of URIs that are archived, the number and
frequency of different timestamped versions of each archived URI,
and the overlap between the archives.  From this, we extrapolate 
the percentage of the surface web that is archived.

%% file: RelatedWork.tex
%


\section{Related Work}

Although the need for Web archiving has been understood since nearly the dawn of the Web \cite{casey-crl98}, these efforts are for the most part independent in motivation, requirements, and scope.  
The Internet Archive\footnote{\url{http://www.archive.org}}, the first archive to attempt global scope, came into existence in 1995 \cite{masanes-book06}.  Since then, many other archives have come into existance.

Recognizing the need for coordination, McCown and Nelson proposed the Web Repository Description Framework and API \cite{mccown-jcdl09}.  Still, there is much to be accomplished before the state of Web archiving is understood and a coordinated effort.

Although much as been written on the technical, social, legal, and political issues of Web archiving,  little research has been conducted on the archive coverage provided by the existing archives.
Day \cite{day-ecdl03} surveyed a large number of archives while investigating the methods and issues associated with archiving.  Day however does not address coverage.
Thelwall touches on coverage when he addresses international bias in the Internet Archive \cite{thelwall-lisr04}, but does not directly address the percent of the Web that is covered.
McCown and Nelson do address coverage \cite{mccown-ist07}, but their research is limited to search engines caches.

Another aspect of coverage still in open discussion is determining what should be covered.  When Gomes, Freitas, et al. addressed the design of a national Web archive \cite{gomes-design06}, incompleteness was inherent in their compromise design.
Mason argues the Web and digital culture has changed our sense of permanence, which in turn has changed the collecting practices the National Library of New Zealand. \cite{mason-lt56}.
Phillips systematically address the question ``what should be archived?'', but concludes that consensus has not been reached and that the Web's huge volume puts complete archiving out of reach.

Another aspect of Web archives that remained resolved until recently was lack of a standard API.  Van de Sompel, et al. have addressed this with Memento \cite{vandesompel-arxiv09, vandesompel-ldow10}.  Memento is an HTTP-based framework that bridges Web archives with current resources; it provides a standard API for identifying and dereferencing archived resources through datatime negotiation.  In Memento, each original resource (identified with ``URI-R'') has 0 or more archived representations (identified with $\text{URI-M}_{i}$ $(i=1..n)$) that encapsulate the URI-R's state at times $t_{i}$.  Using the Memento API, clients are able to request $\text{URI-M}_{i}$ for a specified URI-R.
Memento is now an IETF Internet Draft \cite{draft-vandesompel-memento}.

%% file: Experiment-Overview.tex
\section{Experiment}
From late November 2010 through early January 2011, we performed an experiment to estimate the percentage of URIs that are represented in Web archives. The primary purpose of this experiment was to estimate the percentage of all publicly-visible URIs that have archive copies available in public archives such as the Internet Archive.  A secondary purpose was to evaluate the quality of Web archiving.  This experiment was accomplished in three parts:
\squishlist
  \item Selecting a sample set of URIs that are representative of the Web as a whole,
  \item Determining the current state of the sample URIs (URI http status, SE index, and estimated age), and
  \item Discovering Mementos for the sample URIs.
\squishend

%% file: Experiment-SampleSelection-intro.tex
%


\subsection{Sample URI Selection}\label{ss:sampleUriSelection}

Discovering every URI that exists is impossible; representative sampling is required.  In previous research several methods were used to select a sample of URIs that are representative of the Web as a whole.  This experiment's sampling followed several of these approaches.  Like many other research projects, we used the Open Directory Project (DMOZ).  We also sampled from search engines using the code and procedures provided by Bar-Yosef \cite{baryossef-jota08} and sampled the Delicious Recent bookmark list and Bitly.  The reasoning behind these sources and the methods used for URI selection are detailed below.

For practical reasons (e.g., search engine query limits and execution
time) we selected a sample size of 1,000 URIs for each of the four
sources.  Table \ref{tab:samplesizes} shows the mean number
of mementos for each of the 1000 URI-Rs (both with and without
URI-Rs with zero mementos) in each sample set, the standard deviation,
and the standard error at a 95\% confidence level.

\begin{table}
\centering
\caption{The four sample sets, all with $n=1000$}
\label{tab:samplesizes}
\begin{tabular}{l|c|c|c}
\hline
\textbf{Collection} & \textbf{Mean} & \textbf{SD} & \textbf{SE} \\
\hline
DMOZ (URI-M>0) & 62.68 & 123.86 & 7.68 \\
DMOZ (all) & 56.85 & 119.35 & 7.40 \\
Delicious (URI-M>0) & 81.44 & 232.02 & 14.38 \\
Delicious (all) & 79.40 & 229.45 & 14.38\\
Bitly (URI-M>0) & 41.64 & 229.18 & 14.20 \\
Bitly (all) & 14.66 & 137.30 & 14.20 \\
SE (URI-M>0) & 6.99 & 25.40 & 1.57 \\
SE (all) & 5.40 & 22.55 & 1.40 \\
\hline
\end{tabular}
\end{table} 

%% file: Experiment-SampleSelection-dmoz.tex
%


\subsubsection{Open Directory Project (DMOZ) Sampling}

Using the Open Directory Project (DMOZ)\footnote{\url{http://www.dmoz.org}} as a URI sample source has a long history \cite{monroe-hicss02, gulli-www05, olston-www08, gulli-www05}.  Although it is an imperfect source for many reasons (e.g. its contents appear to be driven by commercial motives and are likely biased in favor of commercial sites), DMOZ was included because it provides for comparability with previous studies and because it is one of the oldest sources available.  In particular, DMOZ archives dating back to 2000 are readily available, which makes DMOZ a reliable source for old URIs that may no longer exist.

Our URI selection from DMOZ differs from previous methods such as Gulli and Signorini \cite{gulli-www05} in that we used the entire available DMOZ history instead of a single snapshot in time.  In particular, we extracted URIs from every DMOZ archive available\footnote{\url{http://rdf.dmoz.org/rdf/archive}}, which includes 100 snapshots of DMOZ made from July 20, 2000 through October 3, 2010.  First, a combined list of all unique URIs was produced by merging the 100 archives.  During this process, the date of the DMOZ archive in which each URI first existed was captured.  This date is later used as indirect evidence of the URI's creation date.  From this combined list, 3,806 invalid URIs (URIs not in compliance with RFC 3986) were excluded, 1,807 non-HTTP URIs were excluded, and 17,681 URIs with character set encoding errors were excluded.  This resulted in 9,415,486 unique, valid URIs from which to sample.  Note that URIs in the DMOZ sample are biased because Internet Archive uses the DMOZ directory as a seed for site-crawling \cite{ia-faq1}.

%% file: Experiment-SampleSelection-delicious.tex
%

\subsubsection{Delicious Sampling}
The next source for URIs is social bookmarking site. In this paper, sampling from Delicious\footnote{\url{http://www.delicious.com}} was used. Delicious is a social bookmarking service started in 2003; it allows users to tag, save, manage and share web pages from a centralized source.  Delicious provides two main types of bookmarks.  Delicious recent bookmarks are the URIs that have been recently added by users.  Delicious popular bookmarks are the currently most popular bookmarks in the Delicious bookmarks set.  In our experiment, we retrieved 1,000 URIs from the Delicious Recent Random URI Generator\footnote{\url{http://www.delicious.com/recent/?random}} (as of Nov. 22, 2010).  We also considered the Delicious Popular Random URI Generator.  However, it's small set of distinct URIs which didn't provide a good sample. 

%% file: Experiment-SampleSelection-bitly.tex
%

\subsubsection{Bitly Sampling}

The Bitly\footnote{\url{http://bit.ly}} project is a web-based service for URI shortening. 
Its popularity grew as a result of being the default URI 
shortening service on the microblogging service Twitter (from 2009-2010), 
and now enjoys a significant user base of its own. 
Any link posted on Twitter is automatically shortened and reposted. Bitly creates a ``short'' URI that when dereferenced issues an HTTP 301 
redirect to a target URI.  The shortened URI consists of a hash value
of up to six alphanumeric characters appended to \url{http://bit.ly/}, 
for example the hash value \url{A} produces:

\begin{verbatim}
% curl -I http://bit.ly/A
HTTP/1.1 301 Moved
Date: Sun, 30 Jan 2011 16:00:48 GMT
Server: nginx
Location: http://www.wieistmeineip.de/ip-address
...
\end{verbatim}

The shortened URI consumes fewer characters in an SMS message (e.g., a 
Tweet), protects long URIs with arguments and encodings from being
mangled in emails and other contexts, and provides an entry point 
for tracking clicks by appending a ``+'' to the URI: 
\url{http://bit.ly/A+}.  This tracking page reveals when the short URI was 
created, as well as the dereferences and associated contexts for the
dereferences.  The creation time of the Bitly is assumed to be greater 
than or equal to the creation time of the target URI to which the Bitly
redirects.

To sample Bitly, we randomly created a series candidate hash values,
dereferenced the corresponding Bitly URI, and recorded the target URIs
(i.e., the URI in the 
\textit{Location:} response header).  The first 1000 bitlys that 
returned HTTP 301 responses were used.  We also recorded the creation 
time of the Bitlys via their associated ``\url{+}'' pages.


%% file: Experiment-SampleSelection-searchengine.tex
%

\subsubsection{Search Engine Sampling}

Search engines play an important role in web page discovery for
most casual users of the Web. Previous studies have examined the
relationship between the Web as a whole and the portion indexed by
search engines.  A search engine sample should be an excellent
representation of the Web as a whole.  However, the sample
must be random, representative, and unbiased.  One way to tackle
the randomness of this sample is by providing the search
engines with multiple random queries, getting the results
and choosing again at random from them.  This intuitive
approach is feasible but suffers from several deficiencies
and is extremely biased.  The deficiencies reside in the
necessity of creating a completely diverse query list of
all topics and keywords.  Also search engines are normally limited
to providing only about the first 1,000 results.  Bias, on
other hand, comes from the fact that search engines present
results with preference to their page rank.  The higher the
popularity of a certain URI, and its adherence to the query,
the more probable it will appear first in the returned
results.

It is necessary to sample the search engine index efficiently,
at random, covering most aspects, while also removing ranking bias
and popularity completely. Several studies have
investigated solving different aspects of this problem. The
most suitable solution was presented by Bar-Yossef
and Gurevich \cite{baryossef-jota08}.

As illustrated in the Bar-Yossef paper, there are two methods to
implement this unbiased random URL sampler from search engine's
index. The first is by utilizing a pool of phrases to assemble queries that will be
later fed to the search engine. The other approach  is based
on random walks and does not need a preparation step. The first method
was utilized in this paper with a small modification to the first phase
of pool preparation.

Originally a huge corpus should be assembled, from which the
query pool will be created. Instead, the N-grams query list \cite{google-n-grams}
obtained from Google was utilized, and a size of 5 grams were chosen.
A total query pool of 1,176,470,663 queries was collected.
A random sampling of the queries was provided to the URI sampler as
the second phase. A huge number of URIs were produced, and
1,000 were filtered at random to be utilized as the unbiased, random
and representative sample of the indexed web. 

%% file: Experiment-CurrentState.tex
%


\subsection{Current URI State Determination}
\label{ss:currentState}
\begin{table*}[htbp]
\label{table:alarms}
\centering
  \begin{minipage}{8.5cm}
\caption{Sample URIs status on the current Web.}
\label{tab:currentWebStatus}
\begin{tabular}{  l c c c c}
  \hline

\textbf{Status}   &\textbf{DMOZ} &  \textbf{Delicious} & \textbf{Bitly} & \textbf{SE}	\\ \hline
  	200	& 507	&	958			&	488			&	943			\\
	3xx$\Rightarrow$200
							&	192		&		27		&		243		&	17			 \\
	3xx$\Rightarrow$Other
							& 50		&		1		&		36		&	3				 \\
							
	4xx	& 135		&		8		&		197		&	16			\\
	
	5xx	&	4			&		3		&		6		&		0		 \\
	Timeout
							&112		&		3		&		30		&	21			 \\ \hline
\end{tabular}
  \end{minipage}
  \begin{minipage}{8.5cm}
\caption{Number of mementos per URI.}
\label{tab:memPerUri}
\begin{tabular}{  l   c c  c  c }
	\hline
\textbf{Mementos per URI}	&	\textbf{DMOZ}	&	\textbf{Delicious}	 &	\textbf{Bitly}	&	\textbf{SE}			\\ \hline
0 (Not archived)			&	93		&	25				&	648		&	225									 \\
1 										&	46		&	79				&	100		 &	336									\\
2	to 5 								&	142		&	491				&	171		 &	320									\\
6 to 10								&	85		&	35				&	17		 &	35									\\
More than 10 					&	634		&	370				&	64		&	 84									\\ \hline
\end{tabular}
  \end{minipage}

\end{table*}
After selecting the different sample sets of URIs, the current state of each 
URI on the web was determined using four components: the current existence 
(or non-existence) of the resource and the current index of the URI in three 
search engines: Google, Bing, and Yahoo!.

First, the current status of each URI was tested using the \texttt{curl} 
command. The status was classified into one of six categories based on 
the HTTP response code. It was important to divide the 3xx responses into two categories because a 
search engine could carry the redirect information and report the original 
URI as indexed and display the new URI. Table \ref{tab:currentWebStatus} lists 
the results of the current status of the URIs on the web. 95\% of Delicious 
and search engine sample URIs sets are live URIs (status code 200), but only 
50\% of DMOZ and Bitly sample URIs sets are live. The reason is that the 
Delicious sample came from the ``Recent'' added bookmarks which means these URIs 
are still alive and some users are still tagging them. Search engines purge their 
caches soon after they discover URIs that return a 404 response \cite{mccown-widm06}.

Second, we tested the status of each sample URI to see if it was indexed 
by each search engine.
For both Bing and Yahoo search engines, Bing APIs\footnote{\url{http://www.bing.com/developers}} 
and Yahoo BOSS APIs\footnote{\url{http://developer.yahoo.com/search/boss/}}, 
respectively were used to find the URI in these search engine indexes. Note that the APIs 
return different results than those from the web interfaces \cite{mccown-jcdl07}. Google 
indexing task depends on Google Research Search Program APIs\footnote{\url{http://www.google.com/research/university/search/}}. 
In this experiment, a significant difference in the indexed results between Google web 
interface and Google Research Search Program APIs was found. Table \ref{tab:currentSEIndex} 
lists the number of the URIs indexed by the different search engines. Google indexed status 
is shown on two rows; the first one is the number discovered by the API, and the second one 
is the union between the URIs discovered by the APIs and the URIs found by the Memento proxy 
on the Google cache.
\begin{table}
\centering
\caption{Sample URIs status on the search engines.}
\label{tab:currentSEIndex}
\begin{tabular}{  l   c c   c c }
  \hline
  &\textbf{DMOZ} &  \textbf{Delicious} & \textbf{Bitly} & \textbf{SE}	\\ \hline

	Bing 				&	495	&		953		&		218		&	552	\\
	Yahoo  			&	410	&		862		&		225		&	979  \\
	Google\tiny(APIs Only)\normalsize	&	307	&		883		&		243		&	702	\\
	Google\tiny(APIs+Cache)\normalsize& 545	&		951		&		305 	&	732	\\ \hline

\end{tabular}
\end{table}

%% file: Experiment-MementoDiscovery.tex
%


\subsection{Memento Discovery}
\label{ss:discoveringMementos}

Concurrently with determining current status, memento discovery was conducted for each URI in the sample sets. In the memento discovery task, the Memento project's archive proxies and aggregator \cite{vandesompel-arxiv09} were used to get a list of all the accessible, public archived versions (mementos) for each URI. Memento's\footnote{\url{http://www.mementoweb.org/}} approach is based on a straightforward extension of ``Hypertext Transfer Protocol (HTTP)'' that results in a way to seamlessly navigate current and prior versions of web resources which might be held by web archives. For any URI, Memento's \emph{TimeBundle} provides an aggregation of all mementos available from an archive. The ODU Computer Science Memento Aggregator implementation was used to retrieve URI TimeBundles, which searched a large number of archives through archive-specific proxies. For each URI, the proxy queries the archive for the available Mementos for the URI, and returns to the aggregator a list of mementos. The aggregator merges the results from all proxies, then sorts it by date and returns a Timemap for the URI.  Table \ref{tab:archives} describes a list of the archives that used in this task.

\begin{table*}
\centering
\caption{Memento's aggregator proxies list.}
\label{tab:archives}
\begin{tabular}{   p{3.7cm} l p{8.5cm} }
  \hline
\textbf{Archive name}			&		\textbf{URI}		& \textbf{Description}	\\	\hline
Internet Archive 			&	http://www.archive.org 				&	Internet Archive is a non-profit that was founded on 1996 to build an Internet library. Its purposes include offering permanent access to historical collections that exist in digital format	\\ \hline
Google						&	http://www.google.com				& 	Google is a search engine provides a cached version.	\\ \hline
Yahoo						& 	http://www.yahoo.com				&	Yahoo is a search engine provides a cached version.		\\ \hline
Bing						& 	http://www.bing.com					& 	Bing is a search engine provides a cached version.		\\ \hline
Archive-It					&	http://archive-it.org				& 	 Archive-It is a subscription service that allows institutions to build and preserve collections of digital content		\\ \hline
The National Archives		&	http://nationalarchives.gov.uk		&	The National Archives is the UK government's official archive.		\\ \hline
National Archives and Records Administration
							&	http://www.archive.gov				& 	National Archives and Records Administration is the record keeper of all documents and materials created during the course of business conducted by the United States		 Federal Government.\\ \hline
UK Web Archive				&	http://www.webarchive.org.uk		& 	UK Web Archive contains UK websites that publish research, that reflect the diversity of lives, interests and activities throughout the UK, and demonstrate web innovation		\\ \hline
Web Cite					&	http://www.webcitation.org			&	Web Cite is an on-demand archiving system for web-references, which can be used by authors, editors, and publishers of scholarly papers and books, to ensure that cited web-material will remain available to readers in the future.		\\ \hline
ArchiefWeb					&	http://archiefweb.eu				&	 ArchiefWeb is a commercial subscription service that archives websites		\\ \hline
California Digital Library	&	http://webarchives.cdlib.org				&	 Focused cultural archives sponsored by the CDL.\\ \hline
Diigo						&	http://www.diigo.com				& 	Diigo is a social bookmarking site that provides archiving services; subsumed Furl in 2009.		 \\ \hline
\end{tabular}
\end{table*}

%% file: Experiment-AgeEstimation.tex
%


\subsection{URI Age Estimation}
Intuition is that the longer a URI has been available on the Web, the
mementos it will have.  Unfortunately, the actual
creation date of a URI is almost always unavailable\footnote{For a
discussion, see: \url{http://ws-dl.blogspot.com/2010/11/2010-11-05-memento-datetime-is-not-last.html}}.  So to estimate the age of
the URI, we estimate the creation date with the earliest value of: the
date of the first memento, the date of first DMOZ archive that reported
the URI, and the first date the URI was added to Delicious.

%% file: Results.tex
%


\begin{figure*}[h!t]
\def\subfigtopskip{0pt}
\def\subfigcapskip{0pt}
\subfigure[DMOZ]{
\includegraphics[width=3.5in]{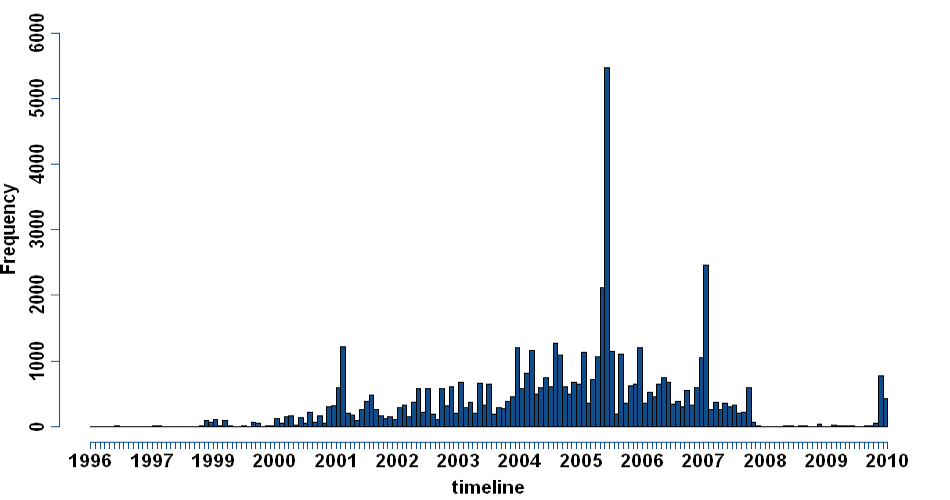}
\label{fig:subfig11}
}
\subfigure[Delicious]{
\includegraphics[width=3.5in]{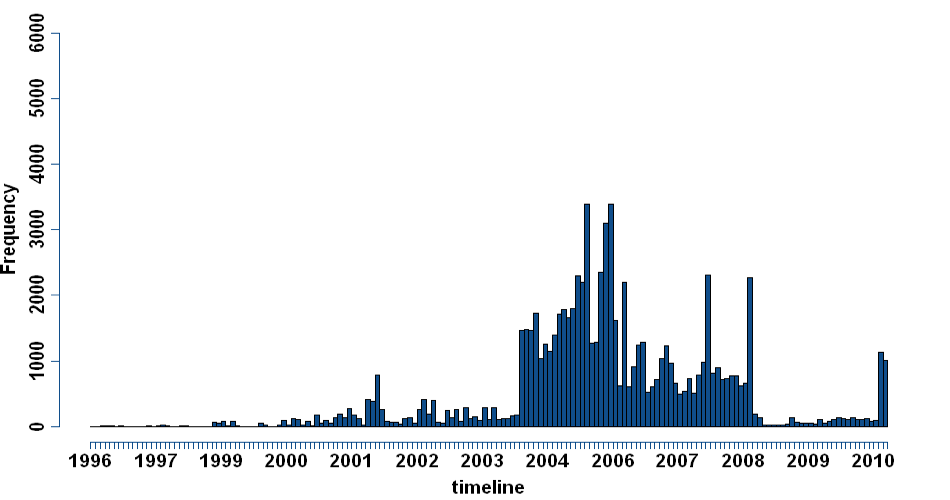}
\label{fig:subfig311}
}
\subfigure[Bitly]{
\includegraphics[width=3.5in]{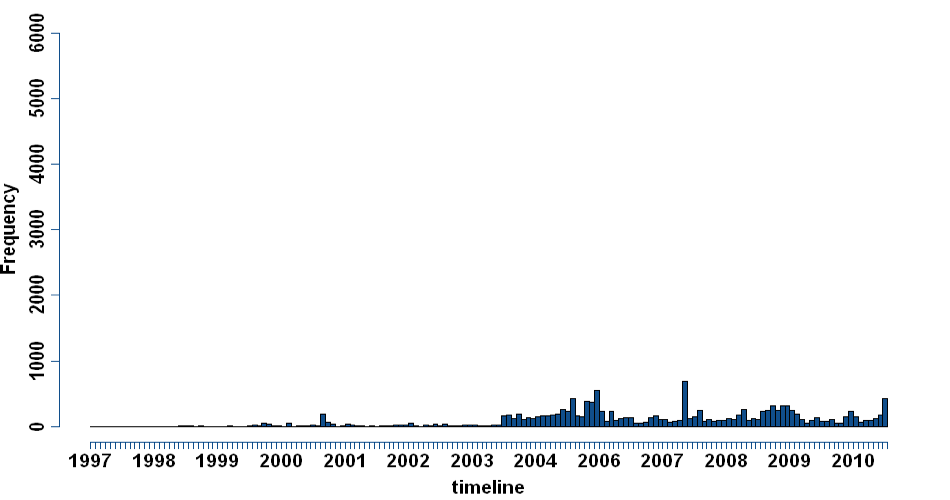}
\label{fig:subfig311}
}
\subfigure[Search Engines]{
\includegraphics[width=3.5in]{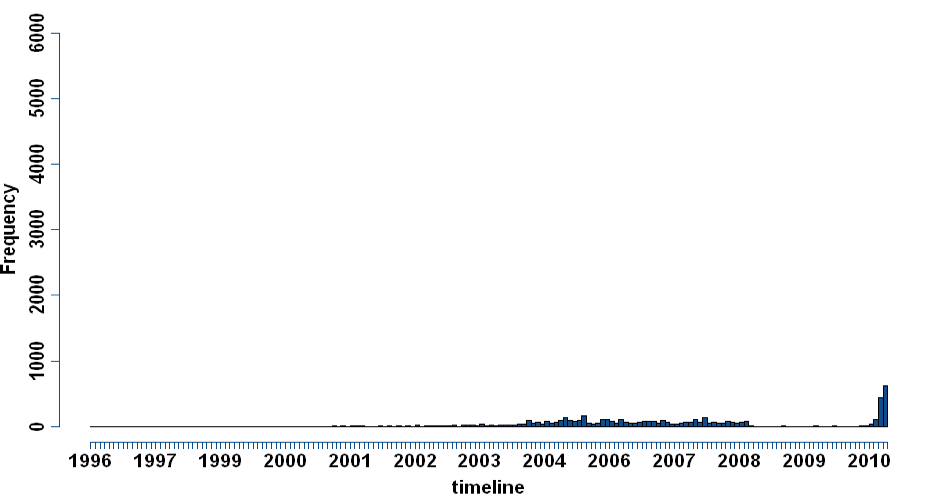}
\label{fig:subfig311}
}
\caption{URI-Ms per month.}
\label{fig:histogram}
\end{figure*}

\begin{figure*}[h!t]
\def\subfigtopskip{0pt}
\def\subfigcapskip{0pt}
\def\subfigbotskip{0pt}
\subfigure[DMOZ]{
\includegraphics[width=3.5in]{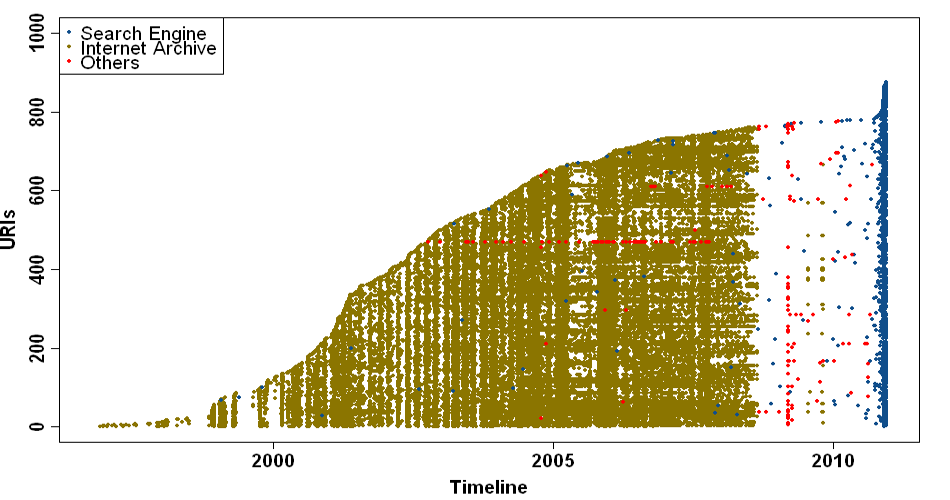}
\label{fig:subfig1}
}
\subfigure[Delicious]{
\includegraphics[width=3.5in]{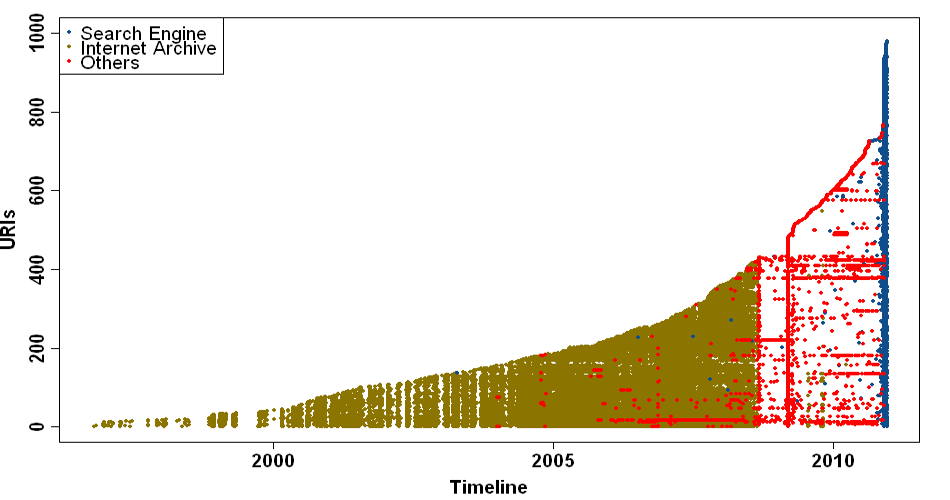}
\label{fig:subfig333}
}
\subfigure[Bitly]{
\includegraphics[width=3.5in]{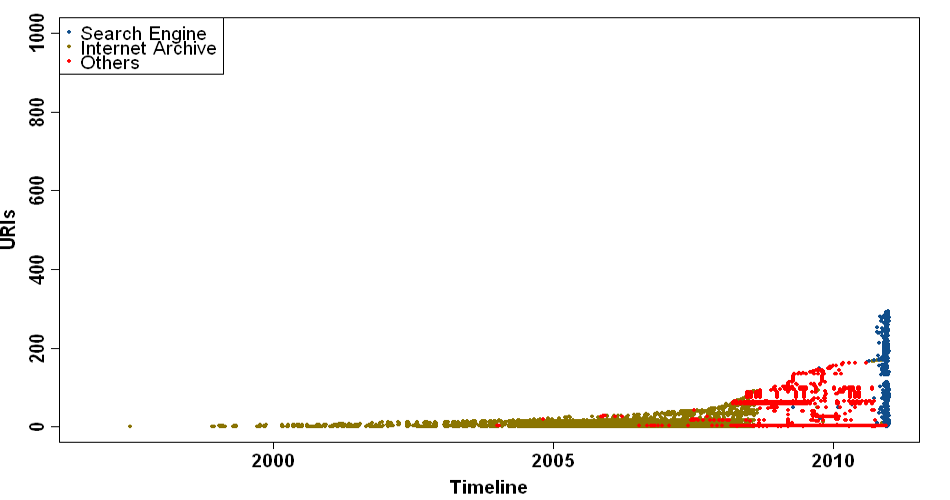}
\label{fig:subfig333}
}
\subfigure[Search Engines]{
\includegraphics[width=3.5in]{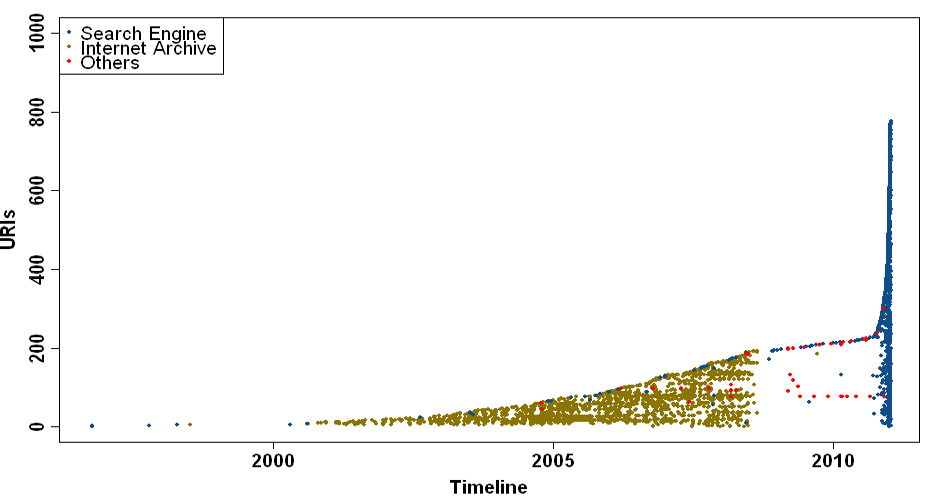}
\label{fig:subfig333}
}
\caption{URI-M distribution.}
\label{fig:scattered}
\end{figure*}

\begin{figure*}[h!t]
\def\subfigtopskip{0pt}
\def\subfigcapskip{0pt}
\def\subfigbotskip{0pt}
\subfigure[DMOZ]{
\includegraphics[width=3.5in]{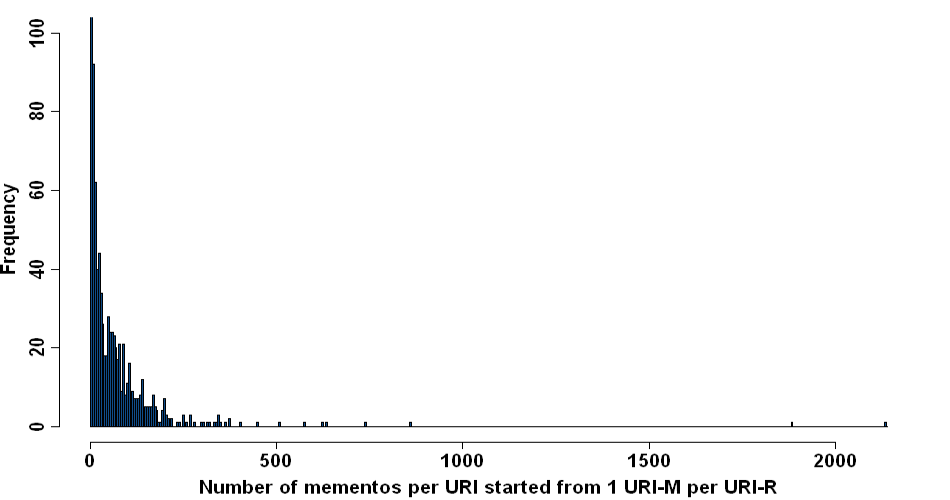}
\label{fig:subfig11}
}
\subfigure[Delicious]{
\includegraphics[width=3.5in]{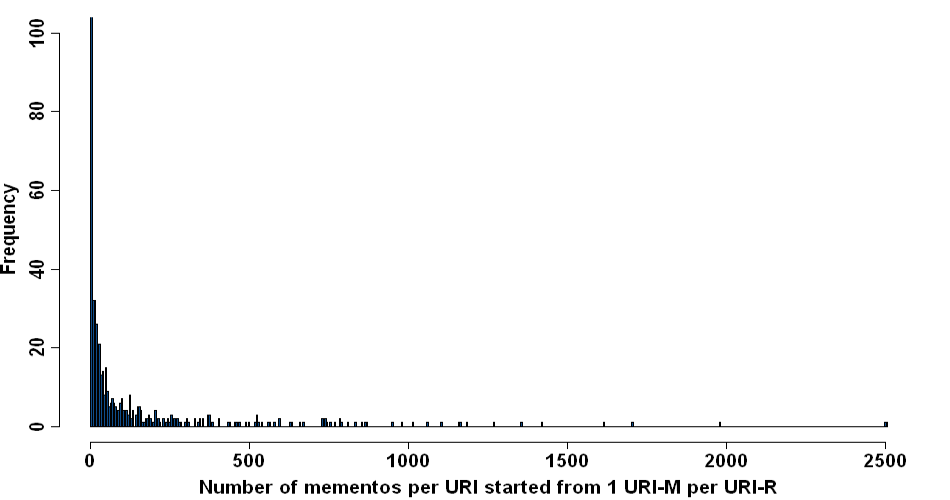}
\label{fig:subfig311}
}
\subfigure[Bitly]{
\includegraphics[width=3.5in]{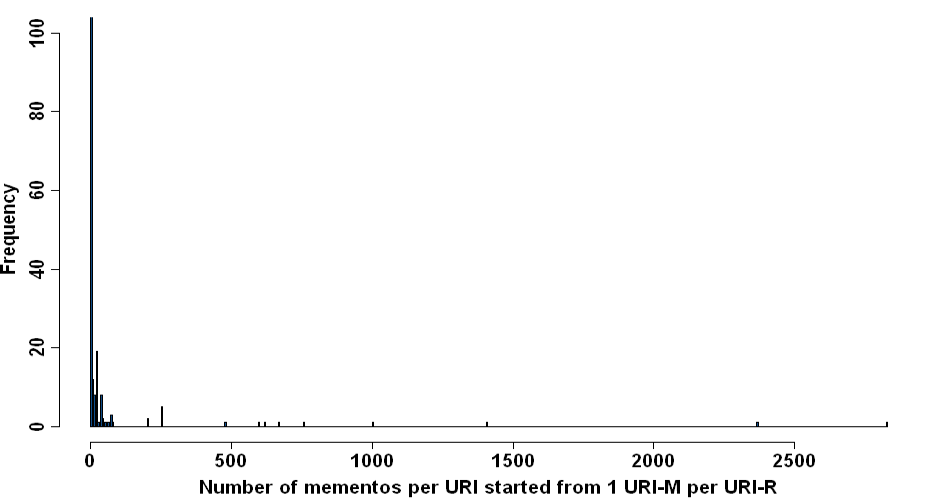}
\label{fig:subfig311}
}
\subfigure[Search Engines]{
\includegraphics[width=3.5in]{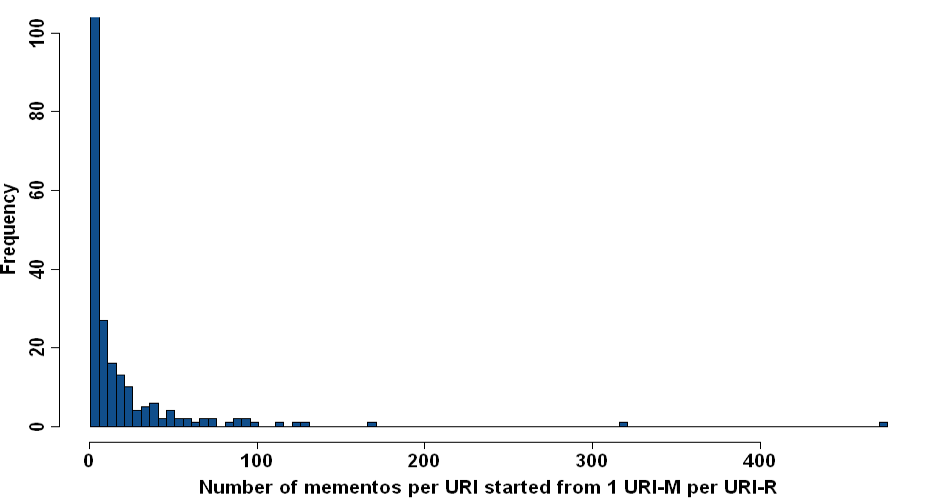}
\label{fig:subfig311}
}
\caption{URI-Ms per URI-R.}
\label{fig:memPerURIHisto}
\end{figure*}

\begin{figure*}[h!t]
\def\subfigtopskip{0pt}
\def\subfigcapskip{0pt}
\def\subfigbotskip{0pt}
\subfigure[DMOZ (14.6\% $\geq$ 1 URI-M/month)]{
\includegraphics[width=3.5in]{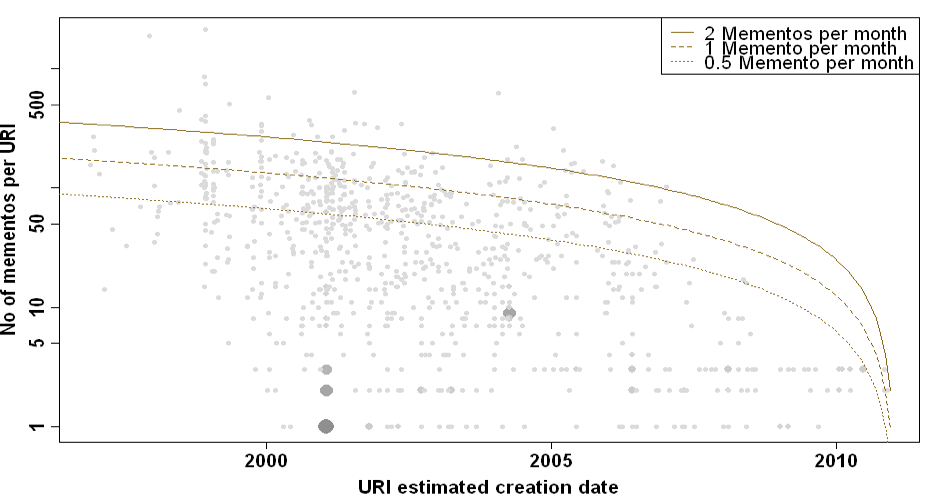}
\label{fig:subfig11}
}
\subfigure[Delicious (31\% $\geq$ 1 URI\-M/month)]{
\includegraphics[width=3.5in]{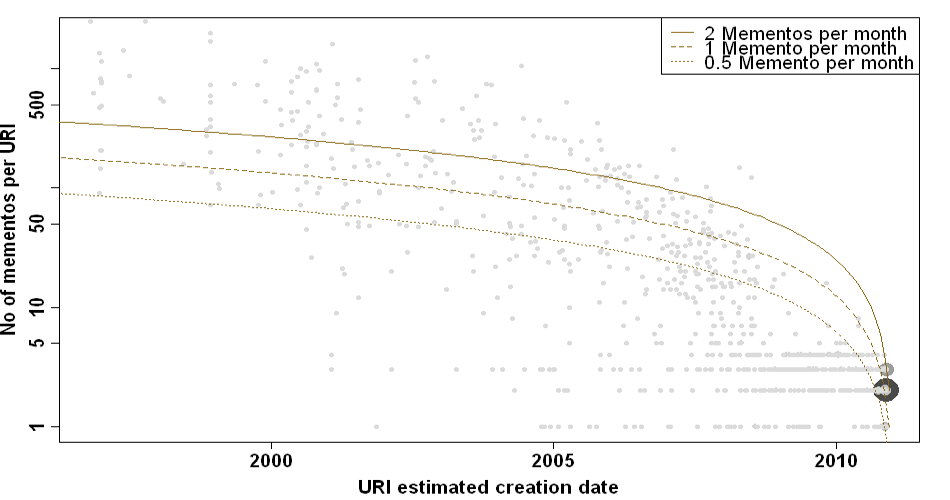}
\label{fig:subfig311}
}
\subfigure[Bitly (14.7\% $\geq$ 1 URI-M/month)]{
\includegraphics[width=3.5in]{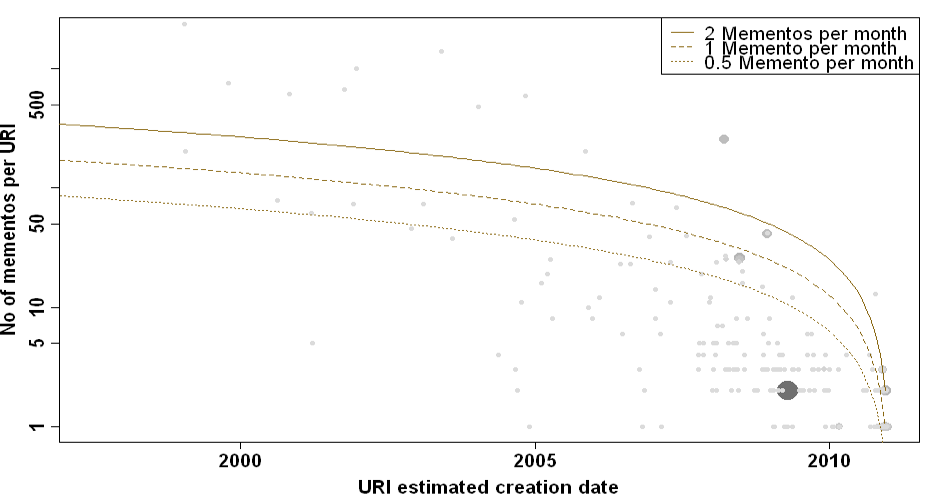}
\label{fig:subfig311}
}
\subfigure[Search Engines (28.7\% $\geq$ 1 URI\-M/month)]{
\includegraphics[width=3.5in]{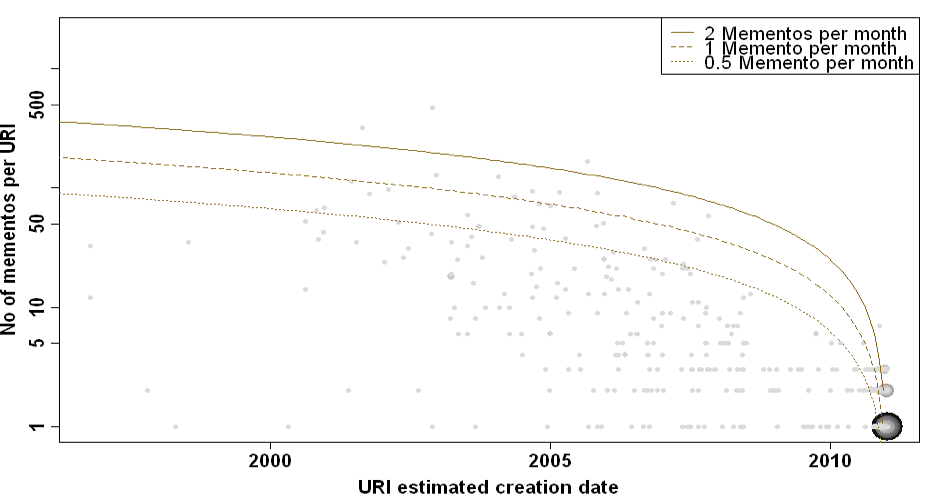}
\label{fig:subfig311}
}
\caption{URI-M density to URI-R age (months).}
\label{fig:densityAge}
\end{figure*}

\section{Results}
The results in this section will be presented by sample set. Figure \ref{fig:histogram} is a histogram of the number of memento URIs retrieved by the Memento proxy servers. This figure shows the distribution of the mementos through time. The DMOZ and Delicious samples have a similar distribution. This figure also shows that there is a low coverage for the period before 2000, as could be expected. The end of 2010 has a good coverage led by the search engine caches.

Figure \ref{fig:scattered} is a detailed graph of the distribution of the mementos over the time.  Each dot represents a single memento.  Dot color indicates the source of the memento: Internet Archive, search engine caches, and other archives.  Reach row on the y-axis represents a single URI-R, which are ordered bottom-to-top by estimated creation date.  The x-axis is the memento's datetime.  Figure \ref{fig:scattered} shows that most of the mementos before 2008 are provided by the Internet Archive. Search engine caches provide very recent copies of the URI.

Table \ref{tab:memPerUri} shows the number of mementos per URI.  The Bitly sample has many URIs that are not covered at all.  This means that Bitly URIs have not been discovered by the various archive's crawlers.  This matches the observation in Table \ref{tab:currentSEIndex} of a poor coverage of the Bitly URIs by the search engine indexes.  The DMOZ sample has many URIs with more than 10 mementos.  There are two reasons for this.  First, DMOZ is a primary input for the Internet Archive's crawler; and second, the DMOZ sample is retrieved from historical DMOZ archives which means these URIs may have existed a long time. Figure \ref{fig:memPerURIHisto} shows the histogram for the number of mementos per URI-R.

Figure \ref{fig:densityAge} shows the density of mementos per datetime.  The x-axis represents the estimated creation date of the URI.  The y-axis represents the number of mementos for a URI-R.  The figure is supplemented with three density guidelines for 0.5, 1 and 2 and mementos per month.

Table \ref{tab:backLink} shows the correlation between the number of mementos and the number of backlinks.  Backlinks are the incoming links to any URI.  The number of backlinks is one indication of the popularity or importance of a website or page.  The number of backlinks was calculated using the Yahoo Boss APIs.  The correlation was calculated using Kendall's $\tau$ and showed a weak positive relationship between the number of backlinks, and the number of mementos.

Table \ref{tab:MementosCoverage} contains statistics about the retrieved mementos by source.  The table reports the number of URI-Ms from the source, the number of URI-Rs covered by the source, the mean URI-Ms per URI-R and corresponding standard deviation (p$\leq$0.01).

\begin{table}
\centering

\caption{Mementos to backlinks correlation ($\tau$).}
\label{tab:backLink}
\begin{tabular}{  l   c c   c p{1.5cm}}
  \hline
  &\textbf{DMOZ} &  \textbf{Delicious} & \textbf{Bitly} & \textbf{SE}	\\ \hline
  	Correlation	& 0.389	&	0.311		&	0.631			&	0.249			\\ \hline
\end{tabular}
\end{table}

\begin{table*}
\centering
\caption{Mementos coverage per type.}
\label{tab:MementosCoverage}
\begin{tabular}{  l  | c c  c  c |c c  c  c }
\hline
	
					&	\multicolumn{4}{c}{\textbf{DMOZ}}			&	 \multicolumn{4}{c}{\textbf{Delicious}}								\\ \cline{2-5} \cline{6-9}
						&	\#URI-M	&	\#URI-R	&	Mean	&	SD			 &	\#URI-M	&	\#URI-R	&	Mean	&	SD					  \\ \hline
Internet Archive		&	55293	&	783	&	70.62	&	130		&	74809	 &	408	&	183.36	&	325												  \\
Google							&	523		&	523	&	1	&	0					 &	897	&	897	&	1	&	0													  \\
Bing								&	427		&	427	&	1	&	0					 &	786	&	786	&	1	&	0												  \\
Yahoo								&	418		&	418	&	1	&	0					 &	479	&	479	&	1	&	0												  \\
Diigo								&	36		&	36	&	1	&	0					 &	354	&	354	&	1	&	0												  \\
Archive-It					&	92		&	4	&	23	&	41				&	 500	&	38	&	13.16	&	30													  \\
National Archives (UK) 	&	25		&	8	&	3.125	&	3		&	521	&	 102	&	5.11	&	10															  \\
NARA					&	5		&	5	&	1	&	0										 &	31	&	19	&	1.63	&	1											  \\
UK Web Archive		&	8		&	5	&	1.6	&	1						&	 391	&	38	&	10.29	&	16																  \\
Web Cite				&	26		&	5	&	5.2	&	8						 &	594	&	57	&	10.42	&	49															  \\
ArchiefWeb			&		-	&	-	&	-	&	-								 &	22	&	3	&	7.33	&	11													  \\
CDLIB					&		-	&	-	&	-	&	-										 &	20	&	5	&	4	&	4												  \\
\hline                                                      	
\hline

							&	\multicolumn{4}{c}{\textbf{Bitly}}	  				 &	\multicolumn{4}{c}{\textbf{Search Engines}}	  \\ \cline{2-5} \cline{6-9}
							&	\#URI-M	&	\#URI-R	&	Mean	&	SD		 &	\#URI-M	&	\#URI-R	&	Mean	&	SD				  \\ \hline
Internet Archive			&	8947	&	70	&	127.81	&	406	&	4067	 &	170	&	23.92	&	49				  \\
Google						&	253	&	253	&	1	&	0								 &	486	&	486	&	1	&	0				  \\
Bing						&	204	&	204	&	1	&	0									 &	515	&	515	&	1	&	0				  \\
Yahoo						&	87	&	87	&	1	&	0									 &	229	&	229	&	1	&	0				  \\
Diigo						&	61	&	61	&	1	&	0									 &	10	&	10	&	1	&	0				  \\
Archive-It					&	75	&	13	&	5.77	&	8					 &	49	&	12	&	4	&	5				  \\
National Archives (UK)		&	531	&	12	&	44.25	&	145	&	1	&	1	 &	1	&	0				  \\
NARA						&	10	&	2	&	5	&	6										 &	4	&	2	&	2	&	0				  \\
UK Web Archive			&	2892	&	32	&	90.38	&	187		&	9	&	 3	&	3	&	3								\\
Web Cite					&	989	&	58	&	17.05	&	82				&	 -	&	-	&	-	&	-				  \\
ArchiefWeb				&	609	&	1	&	609	&		0						 &	-	&	-	&	-	&	-				  \\
CDLIB						&	-	&	-	&	-	&	-											 &	-	&	-	&	-	&	-				  \\ \hline

\end{tabular}
\end{table*}

%% file: Analysis.tex
%


\section{Analysis}
To answer the question ``How much of the Web is archived?'', the memento sources will be compared on 3 axis:

\squishlist
	\item \textit{Coverage}. How many URI-Rs are archived?
	\item \textit{Depth}. How many mementos are available?
	\item \textit{Age}. How old are the mementos?
\squishend

\subsection{Internet Archive}
The results showed that the Internet Archive (IA) has the best coverage and depth. Also, the IA covers URIs from 1996 to the present. IA has a delay from 6-24 months between the crawling and appearance on the archive web interface \cite{ia-faq103}, but the results showed that this (6-24 months) period may be longer as less than 0.1\% of the IA archived versions appeared after 2008. Moreover, IA provides archived versions for dead URIs (response 404).

In late 2010, Internet Archive launched a new WayBack Machine interface\footnote{\url{http://waybackmachine.org}}. The new interface is supported with a new toolbar on archived pages and a new look of the calendar of page captures. The experiment results were retrieved using the classic WayBack Machine, because the new one is still in the beta version.

\subsection{Search engine caches}
Search engines (Google, Bing, and Yahoo) provide a cached version for most indexed pages \cite{mccown-ist07}.  Search engines provide a good coverage, but are limited to 1 copy per URI. Based on our observations, Google and Bing cached copies are kept for a maximum of one month.  Yahoo provides cached version without date, we used a new technique to estimate the cached version age for Yahoo which may be several years \cite{Alsum2011}.

\subsection{Other archives}
The other archives are mainly special purpose archives. These archives provide good coverage for their own web sites which may be limited to country (UK Web Archive, The National Archives, and NARA), special subscriber collections (Archive-It, ArchiefWeb, and CDLIB), or the user preferences (WebCite, and Diigo). Most of these sites have a high number of archived copies for these URIs. The age for the archived copies depends on the age of the archive itself.

\subsection{Archive overlap}
\begin{figure*}[b!]
	\centering
	\includegraphics[width=0.7\textwidth]{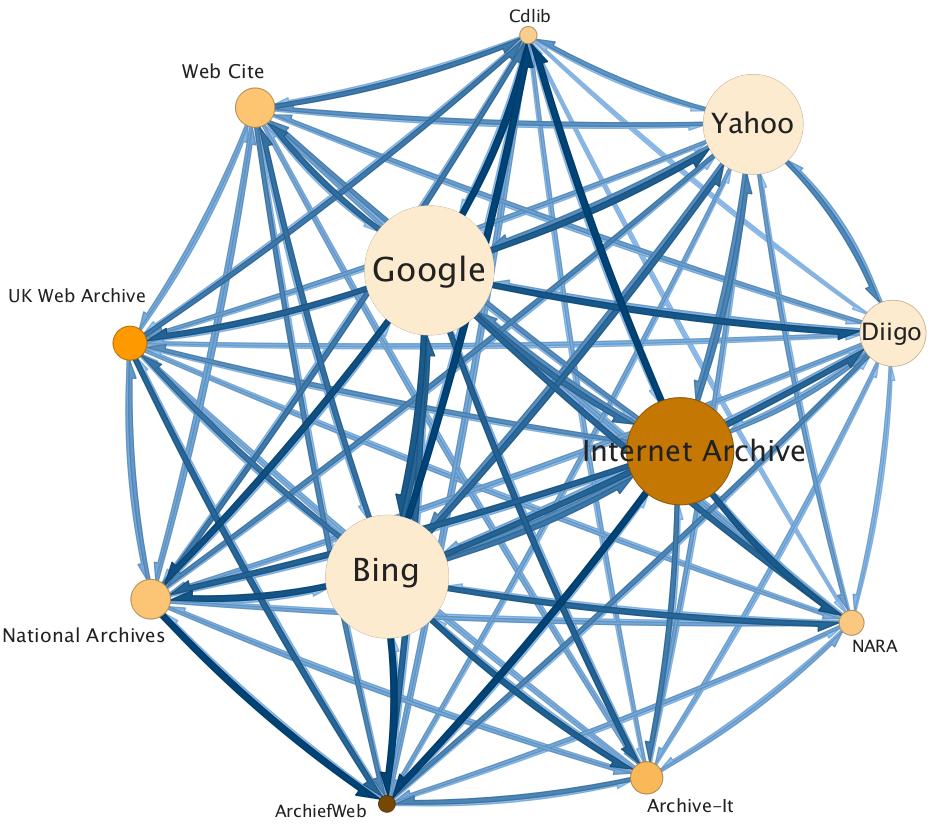}
	\caption{Archive connections graph}
	\label{fig:Mapequation}
\end{figure*}

We also looked at the relationships and overlap between sources, which is shown in
figure \ref{fig:Mapequation}.  The diameter of each circle represents relative coverage.  The color density expresses the depth. The width of the edges expresses the intersection between both archives. The visualization was done by MapEquation\footnote{\url{http://www.mapequation.org}}.


%% file: Conclusion.tex
%


\section{Conclusion and Future work}
When we began the research to answer the question ``How much of the Web is archived?'' we did not anticipate a quick, easy answer.  Suspecting that different URI sample sources would yield different results, we chose sources we believed would be independent from each other and provide enough insight to allow estimation of a reasonable answer.  Instead we have found that the URI source is a significant driver of its archival status.

Consider the graphs in Figure \ref{fig:scattered}.  Clearly, the archival rate (URIs with at least one memento) is much higher for the DMOZ and Delicious samples than for the Bitly and search engine samples, which also differ considerably from each other.  URIs from DMOZ and Delicious have a very high probability of being archived at least once.  On the other hand, URIs from search engine sampling have about $2/3$ chance of being archived and Bitly URIs just under $1/3$.  This leads us to consider the reasons for these differences.

Something that DMOZ and Delicious have in common is that a person actively submits each URI\footnote{Spam-bots notwithstanding} to the service.  (Indeed DMOZ and Delicious can be both considered directory services, with DMOZ representing Web 1.0 and Delicious representing Web 2.0.)  Search engine sample URIs, however, are not submitted in the same way.  The process used by search engines is more passive, with URI discovery depending on search engine crawl heuristics.  Bitly is more of a mystery.  Our research did not delve into how Bitly URIs are used.  But the low archival rate leads us to think that many private, stand-alone, or temporary resources are represented.  These substantial archival rate differences have led us to think that publicity interest a URI receives is a key driver of archival rate for publicly-accessible URIs.

Future work will include study of the relationship between the rate of change of the web page and the rate of the archiving process. Also, we plan to study of the quality of the archive itself. This work has been done on a general sample of URIs. In future work, the archived URIs will be studied based on specific languages beyond English.